\documentclass[aps,prl,twocolumn,superscriptaddress]{revtex4}
\usepackage{amssymb}
\usepackage{bbm}
\usepackage{indentfirst}
\usepackage{graphicx}
\usepackage{amsmath,amssymb}
\usepackage{bpchem}

\usepackage{color}
\usepackage{tabularx}
\usepackage{multirow}
\usepackage{makecell}
\usepackage{array}
\usepackage{appendix}

\makeatletter
\renewcommand{\subsection}{\@startsection{subsection}{2}{\z@}%
	{3.25ex \@plus1ex \@minus.2ex}%
	{1.5ex \@plus.2ex}%
	{\normalfont\footnotesize\bfseries}}
\makeatother

\begin{document}
	
	\title{Quantized Frequency-locking and Extreme Transitions in a Ring of Phase Oscillators with Three-Body Interactions}
	
	\author{Jinfeng Liang}\thanks{These authors contributed equally to this work.}
	\affiliation{School of Physical Science and Technology, Beijing University of Posts and Telecommunications, Beijing, 100876, People's Republic of China}
	\author{Shanshan Zhu}\thanks{These authors contributed equally to this work.}
	\affiliation{School of Physical Science and Technology, Beijing University of Posts and Telecommunications, Beijing, 100876, People's Republic of China}
	\author{Yang Li}\email{yangli2025@hbnu.edu.cn}
	\affiliation{Huangshi Key Laboratory of Metaverse and Virtual Simulation, School of Mathematics and Statistics, Hubei Normal University, Huangshi, 435002, China}
	\author{Qionglin Dai}\email{qldai@bupt.edu.cn}
	\affiliation{School of Physical Science and Technology, Beijing University of Posts and Telecommunications, Beijing, 100876, People's Republic of China}
	\author{Haihong Li}
	\affiliation{School of Physical Science and Technology, Beijing University of Posts and Telecommunications, Beijing, 100876, People's Republic of China}
	\author{Junzhong Yang}\email{jzyang@bupt.edu.cn}
	\affiliation{School of Physical Science and Technology, Beijing University of Posts and Telecommunications, Beijing, 100876, People's Republic of China}

	\begin{abstract}
		We report a spectrum of exotic frequency-locked states in a ring of phase oscillators with pure three-body interactions. For identical oscillators, the system hosts a vast multiplicity of stable quantized frequency-locked states without phase coherence. Introducing frequency heterogeneity broadens each quantized level into a continuous band and drives an extreme second-order transition at $\Delta_c$: below $\Delta_c$ the entire population locks to a collective phase velocity; above $\Delta_c$ a desynchronous state emerges, characterized by strongly localized bursts on a slowly varying background. This minimal model thus establishes a new paradigm for complex synchronization landscapes arising from higher-order interactions.
		
	\end{abstract}
	
	\maketitle
	
	Synchronization is a universal collective phenomenon in which interacting oscillators spontaneously align their rhythms, leading to coordinated behaviour across a system \cite{pik01,stro}. Examples span the natural, social and engineered worlds: fireflies flash in unison \cite{erm91} and cardiac pacemaker cells beat as one \cite{hay17}; Josephson-junction arrays \cite{wie96,wie98}, atomic-recoil lasers \cite{jav08}, and spin-torque oscillator arrays \cite{tur17} exhibit coherent dynamics; audiences fall into rhythmic applause \cite{ned00} and pedestrians into synchronized steps on footbridges \cite{eck07}. Equally crucial are engineered realizations such as power grids, whose stable operation hinges on phase-locked generators \cite{roh12,fil08,wit22}.
	
	The Kuramoto model, an ensemble of phase oscillators with distributed natural frequencies coupled through all-to-all pairwise interactions, provides the canonical description of these phenomena \cite{kur75,str00}. It displays a continuous transition from incoherence to macroscopic synchrony once the coupling exceeds a critical threshold. Over the past decades, this paradigm has been extended along two principal axes: from global coupling to non-local or purely local network topologies \cite{ace05,rod16,kur02}, and from pairwise to higher-order interactions that simultaneously involve three or more oscillators \cite{boc23,iac19,gri17}. These generalizations have revealed a rich bestiary of coherent states, including twisted states \cite{ome14} and chimera states \cite{abr04,zhu14}, as well as exotic transitions such as explosive ones \cite{gom11,boc16,hu14}. Yet even minimal models continue to yield unexpected collective behaviours.
	
	In this letter, we report unexpected collective dynamics in what is arguably the simplest higher-order setting, a ring of $N$ phase oscillators with three-body interactions. We find: i) Quantized frequency-locking without phase coherence. The system with identical oscillators supports a vast multiplicity of quantized frequency-locked states with collective velocities $\sin(2\pi n/N)$ ($n$ the integers). These states possess no macroscopic phase coherence yet are linearly stable; their number grows linearly with the system size; ii) From discrete velocities to continuous bands. As frequency heterogeneity increases, the sharp quantization of collective velocities dissolves into broad, overlapping bands; iii) Localized bursting on a slowly varying background. Strong frequency heterogeneity destroys global locking and gives rise to desynchronous states in which passive units remain close to the previous quantized states while active oscillators execute rapid bursts, yielding a pronounced spatial localization of velocity fluctuations; iv) An extreme continuous transition. For uniformly distributed frequencies, the passage from the desynchronous to the quantized regime is continuous and extreme: the disordered side is fully asynchronous, yet immediately below the critical heterogeneity every oscillator locks to the same collective velocity.
	
	We consider a ring of $N$ phase oscillators, whose dynamics are governed by the following equation
	\begin{equation} \label{eq1}
		\dot{\theta}_{i}= \omega_{i}+K\sin(\theta_{i+1}-2\theta_{i}+\theta_{i-1}),
	\end{equation}
	where $\theta_i$ denotes the phase of oscillator $i$, with site index $i$ taken modulo $N$. The natural frequency $\omega_i$ of each oscillator is drawn from a probability distribution $g(\omega)$. Without loss of generality, we assume that the mean of $g(\omega)$ is zero. In this letter, we consider two types of frequency distributions, a uniform distribution $g(\omega)=1/\Delta$ for $\omega\in(-\Delta/2,\Delta/2)$, and a Lorentzian distribution $g(\omega)=\Delta/\pi(\omega^2+\Delta^2)$. The parameter $\Delta$ characterizes the degree of frequency heterogeneity. In the limit $\Delta=0$, all oscillators are identical. The interaction term $\sin(\theta_{i+1}-2\theta_{i}+\theta_{i-1})$ represents a three-body interaction, where each oscillator interacts with its two nearest neighbors. In its linearized form, the coupling reduces to a standard diffusive interaction. The coupling strength is denoted by $K$. We set $K=1$ without loss of generality, leaving $\Delta$ as the sole control parameter governing the system's behaviour.
	
	\begin{figure*}[t]
		\includegraphics[width=2\columnwidth]{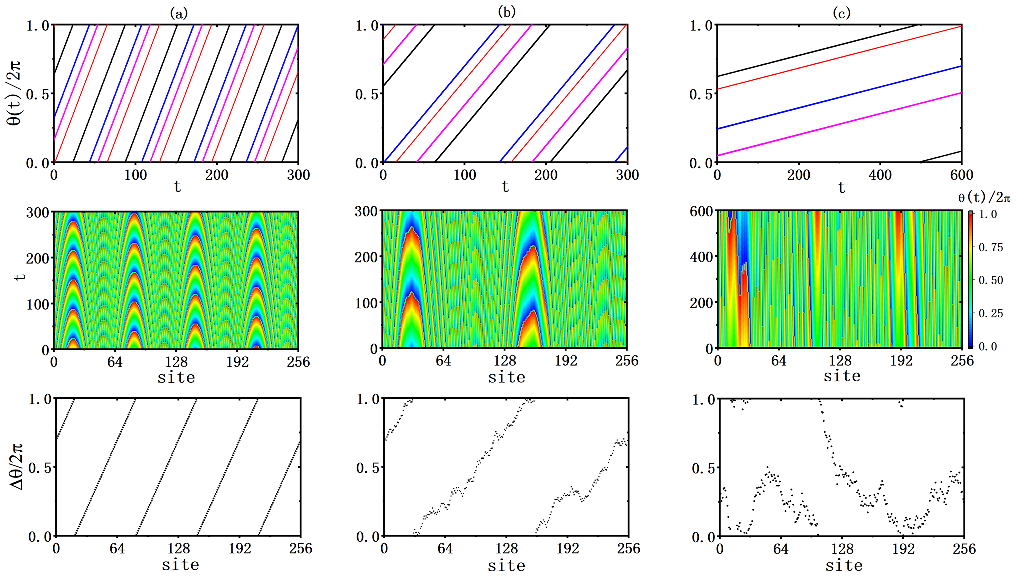}
		\caption{\label{fig1}Illustration of quantized frequency-locked states for increasing frequency heterogeneity: (a) $\Delta=0$, (b) $\Delta=0.4$, and (c) $\Delta=1$. Top panels: phase evolutions of four representative oscillators (color-coded); all rotate at the same constant collective velocity. Middle panels: space-time plot of the oscillator phases; the regular periodic pattern degrades as $\Delta$ grows. Bottom panels: the spatial distribution of local phase gradient $\Delta\theta$, which accumulates $8\pi$ (a), $4\pi$ (b), and $-2\pi$ (c) when traversing the ring from site 1 to 256. Other parameters: $K=1$ and $N=256$.} 
	\end{figure*}
	
	\makeatletter 
	\renewcommand{\section}{\@startsection{section}{1}{0mm}
		{-\baselineskip}{0.5\baselineskip}{\bf\leftline}}
	\makeatother

	\section{Results}
	\subsection{Quantized synchronous states}
	We consider phase oscillators with natural frequencies uniformly distributed. Starting with the case of identical oscillators ($\Delta=0$), we set $N=256$ and numerically simulate model~(\ref{eq1}). A representative example with a random initial condition is shown in Fig.~\ref{fig1}(a). The top panel illustrates the temporal evolution of several representative oscillators, all advancing with the same constant phase velocity $\Omega$, signalling system-wide frequency locking. The spatiotemporal dynamics, depicted in the middle panel, reveals two key features, the absence of global phase coherence and the presence of spatial periodicity in the phase distribution. The lack of global phase coherence is evident from the fact that, at any given time, the oscillator phases are spread over the entire range of [0, $2\pi$). This behaviour contrasts with conventional synchronized states, where the order parameter $Z=\sum_i \exp(j\theta_i)/N$ (with $j$ being the imaginary unit) typically assumes a non-zero value. Meanwhile, the spatial structure of the phase profile exhibits periodic pattern, characterized by alternating regions of rapid and gradual phase variation across the ring. To further quantify this spatial pattern, we define the local phase gradient $\Delta\theta_i=\theta_{i+1}-\theta_i$ for each oscillator $i$. As shown in the bottom panel, $\Delta\theta_i$ increases linearly with the site index. This linear dependence signals the absence of local phase coherence defined by $Z'=\sum_i \exp(j(\theta_i-\theta_{i-1}))/N$, in stark contrast to twist states, which retain local phase coherence even though they lack global phase order. Moreover, $\Delta\theta_i$ traverses $2\pi n$ (with $n=4$) when oscillator crosses the ring, which determines the spatial period of the pattern to be $N/n$.
	
	Upon introducing moderate heterogeneity in natural frequencies, specifically with $\Delta=0.4$ in Fig.~\ref{fig1}(b) and $\Delta=1$ in Fig.~\ref{fig1} (c), the phase oscillators maintain global frequency synchronization. However, the spatial regularity of the phase distribution gradually deteriorates as the frequency heterogeneity increases. For instance, a discernible spatial pattern is still observed at $\Delta=0.4$, whereas it is completely lost at $\Delta=1$. Concomitantly, the phase gradients $\Delta\theta_i$ lose their strictly linear dependence on site index: at $\Delta=0.4$ they may reverse locally, and at $\Delta=1$ they vary highly irregularly. Nevertheless, a global pattern persists, as the oscillator index traverses the ring, the accumulative change of $\Delta\theta_i$ remains quantized to $2\pi n$, with $n=2$ in Fig.~\ref{fig1}(b) and $n=-1$ in Fig.~\ref{fig1}(c).
	
	To determine the collective phase velocity $\Omega$ for the frequency-locked states shown in Fig.~\ref{fig1}, we recast model~(\ref{eq1}) as
	\begin{equation} \label{eq2}
		\dot{\theta}_{i}=\omega_i+\sin(\Delta\theta_{i}-\Delta\theta_{i-1}).
	\end{equation}
	In the steady state, $\theta_i(t)=\Omega t+\phi_{i}$, so
	\begin{eqnarray}\label{eq3}
		\Omega=\omega_i+\sin(\Delta\theta_{i}-\Delta\theta_{i-1}),
	\end{eqnarray}
	which give rises to the increment of local phase gradient $\Delta\theta_{i}-\Delta\theta_{i-1}$ to be either $\sin^{-1}(\Omega-\omega_i)$ or $\pi-\sin^{-1}(\Omega-\omega_i)$.
	
	Real solutions require $|\Omega-\omega_i|\leq 1$ for every oscillator; hence the extremal natural frequencies must satisfy $|\Omega-\omega_{\max}|\leq 1$ and $|\Omega-\omega_{\min}|\leq 1$, which imposes the frequency heterogeneity $\Delta$ the upper bound
	\begin{equation}\label{eq4}
		\Delta\equiv\omega_{\text{max}}-\omega_{\text{min}}\le 2\equiv\Delta_c.
	\end{equation}
	
	The ring structure demands
	\begin{eqnarray}\label{eq6}
		\sum_{i=1}^{N}(\Delta\theta_i-\Delta\theta_{i-1})=2\pi n, \quad n\in \mathbb{Z}.
	\end{eqnarray}
	Then, we have
	\begin{equation}\label{eq7}
		\sum_{i=1}^{N}\sin^{-1}(\Omega-\omega_i)=2\pi n,  \quad n=-\frac{N}{4},\cdots,\frac{N}{4},
	\end{equation}
	which quantizes the admissible velocities $\Omega_n$. For identical oscillators ($\Delta=0$), Eq.~(\ref{eq7}) gives rise to
	\begin{equation}\label{eq8}
		\Omega_n=\sin\left(\frac{2\pi n}{N}\right),
	\end{equation}
	which accurately gives $\Omega_4$ for the phase velocity shown in Fig.~\ref{fig1}(a). Given a sample of natural frequencies $\{\omega_i\}$, Eq.~(\ref{eq7}) allows for a vast multiplicity of frequency-locked states, and for $N\rightarrow\infty$, an infinite family of frequency-locked states coexist in model~(\ref{eq1}). We designate these states as \textit{quantized frequency-locked states} with $\Omega_n$ labelling the level-$n$ state. An extensive family of coexisting synchronous states has been reported for the Kuramoto model augmented by a second-harmonic coupling \cite{tak11,kom13,li14}. There, all stable synchronous states share the same collective phase velocity, and the degeneracy arises from a $\pi$-shift symmetry that identifies $\theta$ with $\theta+\pi$, a mechanism wholly distinct from the quantized, yet coherence-free, frequency-locked states reported here.
	
	To examine the linear stabilities of the quantized frequency-locked states, we recast Eq.~(\ref{eq2}) in terms of the local phase gradients
	\begin{eqnarray} \label{eq9}
		\Delta\dot{\theta}_{i}=\omega_{i+1}&-&\omega_i+\sin(\Delta\theta_{i+1}-\Delta\theta_{i})\nonumber\\
		&-&\sin(\Delta\theta_{i}-\Delta\theta_{i-1}).
	\end{eqnarray}
	Level-$n$ quantized state corresponds to the equilibrium $\Delta\theta_i=\Delta\theta_i^*$ of Eq.~(\ref{eq9}) that satisfies Eq.~(\ref{eq3}) and Eq.~(\ref{eq6}). Let $\delta\Delta\theta_i$ denote a small perturbation to level-$n$ quantized state; linearizing gives
	\begin{eqnarray} \label{eq10}
		\delta\Delta\dot{\theta}_{i}(t)&=&\cos(\Delta\theta_{i+1}^*-\Delta\theta_{i}^*)(\delta\Delta\theta_{i+1}-\delta\Delta\theta_{i})\\
		&-&\cos(\Delta\theta_{i}^*-\Delta\theta_{i-1}^*)(\delta\Delta\theta_{i}-\delta\Delta\theta_{i-1}). \nonumber
	\end{eqnarray}
	The stability of the equilibrium is therefore governed by the Jacobian matrix $\mathbf{DF}$ (Details are given in the Supplemental Material). For identical oscillators ($\Delta=0$) and $\Delta\theta_{i+1}^*-\Delta\theta_{i}^*=2\pi n/N$, $\mathbf{DF}$ becomes the circulant matrix
	\begin{eqnarray} \label{eq11}
		\mathbf{DF}=\cos\frac{2\pi n}{N}\left(
		\begin{array}{cccccc}
			-2 & 1 & 0 & \cdots & 0 & 1 \\
			1 & -2 &  1 & \cdots & 0 & 0 \\
			\vdots & \vdots & \vdots & \vdots & \vdots & \vdots \\
			1 & 0 & 0 & \cdots & 1 & -2 \\
		\end{array}
		\right).
	\end{eqnarray}
	Its eigenvalues are
	\begin{eqnarray}\label{eq12}
		\lambda_k=2\cos\frac{2\pi n}{N}(-1+\cos\frac{2k\pi}{N}),
	\end{eqnarray}
	with $k=0,\cdots,N-1$. Since $|n|\leq N/4$, the spectrum contains a single zero eigenvalue $\lambda_{0}=0$ corresponding to the global phase rotation mode, while all remaining eigenvalues $\lambda_{k>0}$ are strictly negative. Consequently, the quantized frequency-locked states are asymptotically stable against all desynchronizing perturbations.
	
\begin{figure}[t]
		\includegraphics[width=1\columnwidth]{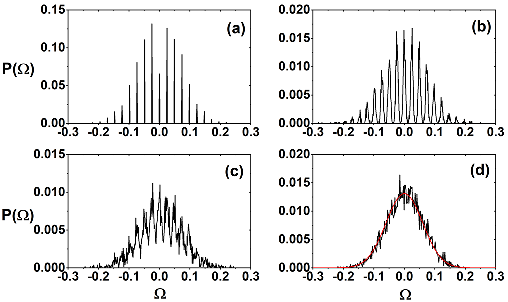}
		\caption{\label{fig2} Empirical probability density $P(\Omega)$ of the collective phase velocity for (a) $\Delta=0$, (b) $\Delta=0.2$, (c) $\Delta=0.4$, and (d) $\Delta=1$. As $\Delta$ increases, $P(\Omega)$ changes from sharp quantized lines (a) to broadened bands (b,c) and finally to a single continuous peak (d); the red curve in (d) is a Gaussian fit.}
	\end{figure}

	Beyond linear stability, basin stability, quantified by the size of the attraction basin, becomes crucial when multiple synchronous states coexist \cite{men13,zhu08}. To estimate the basin stability of each quantized frequency-locked state, we perform 5000 independent simulations starting from random initial phases and random natural-frequency samples, then record the empirical probability density $P(\Omega)$ of the observed collective velocity $\Omega$. Figure~\ref{fig2} traces the evolution of $P(\Omega)$ from sharp $\Omega_n$ peaks to a smooth spectrum with the increase of the frequency heterogeneity. For identical oscillators ($\Delta=0$), $P(\Omega)$ is non-zero only at the predicted quantized values $\Omega_n=\sin(2\pi n/N)$.  The weight $P(\Omega_n)$ decreases monotonically with $|n|$ and vanishes for $|n|>10$, indicating that low-$|n|$ states possess vastly larger basins.  States with $|n|>10$ are never reached from random initial conditions, even though they are linearly stable. Finite frequency heterogeneity changes the picture. Equation~(\ref{eq7}) implies that each nominal level-$n$ state broadens into a continuous band of velocities whose width grows with $\Delta$. When $\Delta$ is sufficiently large, the bands overlap and the discrete hierarchy is lost. In the strong frequency heterogeneity regime, the observed distribution is well approximated by a Gaussian, a feature previously reported for twist states in a ring of phase oscillators \cite{wil06} and for in-phase/anti-phase states in one-dimensional $J_1$-$J_2$ phase oscillator model \cite{chen24}.
	
	\subsection{Desynchronous states}
\begin{figure}[t]
		\includegraphics[width=1\columnwidth]{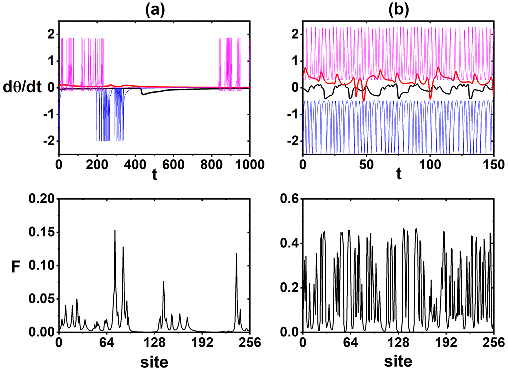}
		\caption{\label{fig3} Desynchronous states for (a) $\Delta=2.04$ and (b) $\Delta=3$. Top panels: instantaneous phase velocities $\dot{\theta}(t)$ of two active oscillators ($|\omega_i|>1$, black and red) and two passive ones ($|\omega_i|<1$, blue and magenta). Passive units remain close to the quantized frequency-locked background, whereas active units exhibit bursting dynamics. Bottom panels: spatial profile of the velocity fluctuation $F_i$. Bursting is strongly localized just above $\Delta_c$ and becomes more dispersed as $\Delta$ increases.}
	\end{figure}

	The quantized frequency-locked states exist only while the frequency heterogeneity remains below the critical value $\Delta_c=2$.  Once $\Delta>\Delta_c$, the system contains oscillators whose natural frequency satisfies $|\omega_i|>1$ (active oscillators); these units destroy global locking and seed desynchronous states. In such states, active and passive ($|\omega_i|<1$) oscillators differ qualitatively in their phase-velocity dynamics. The top panels in Figs.~\ref{fig3}(a) and (b) illustrate the instantaneous phase velocities for $\Delta=2.04$ and $3$, respectively.  Active units exhibit sharp, intermittent bursts that ride on the slow, quantized frequency-locked background. As $\Delta$ grows, the bursts become more frequent and ultimately merge into rapid, regular oscillations. Passive units, by contrast, remain locked to the slow background except when they are immediately adjacent to active sites. We quantify the strength of bursting for oscillator $i$ by the time-averaged fluctuation $F_i=\int_0^T \dot{\theta}_i^2 dt/T$ with $T$ taken large after the transient. Large $F_i$  signals vigorous bursting; small $F_i$ indicates quiescent, nearly frequency-locked motion. The spatial profile of $F_i$ for $\Delta=2.04$ in the bottom panel in Fig.~\ref{fig3}(a) shows pronounced, well-separated peaks that coincide with the active oscillators, embedded in a low-level background contributed by the passive majority. Thus, bursting is strongly localized just above $\Delta_c$. Increasing $\Delta$ adds more active sites, enlarges the number of bursting peaks, and eventually produces overlap that drives the system into a delocalized regime (see Fig.~\ref{fig3}(b)).
	
	\subsection{Synchronization transition}
	The transition between quantized frequency-locked and desynchronous states at $\Delta_c$ is an extreme synchronization transition: on the disordered side all oscillators are asynchronous, while on the ordered side all of them get locked to the same collective phase velocity immediately at $\Delta_c$. Unlike previously reported first-order extreme synchronization transitions \cite{cal20,lee25}, the present one is continuous for any system size. To demonstrate it, we monitor two global observables as functions of the frequency heterogeneity $\Delta$, the spatially averaged fluctuation $\langle F\rangle=\sum_i F_i/N$, which measures how far the system departs from the quantized frequency-locked background, and the inverse participation ratio $\mathrm{IPR}=\sum_i F_i^2/(\sum_i F_i)^2$ \cite{eve00,lon25}, which quantifies the degree of localization of bursting activity. For each $\Delta$, $\langle F\rangle$ and $\mathrm{IPR}$ are ensemble averaged over $50$ realizations with random initial phases and random natural-frequency samples. Figure~\ref{fig4}(a) shows that $\langle F\rangle$ remains vanishingly small for $\Delta<2$ and then grows approximately linearly beyond this value, pinpointing the analytically obtained critical point $\Delta_c=2$. The $\mathrm{IPR}$, in contrast, develops a sharp peak exactly at $\Delta_c$, signalling that bursting is maximally localized at the transition. In the language of second-order ferromagnetic phase transitions, $\langle F\rangle$ plays the role of the order parameter, rising continuously from zero at the critical point, while the $\mathrm{IPR}$ functions as the magnetic susceptibility, exhibiting a sharp peak there. Together, these signatures confirm a continuous extreme-synchronization transition.
	
	\subsection{Lorentzian distribution}
\begin{figure}[t]
		\includegraphics[width=1\columnwidth]{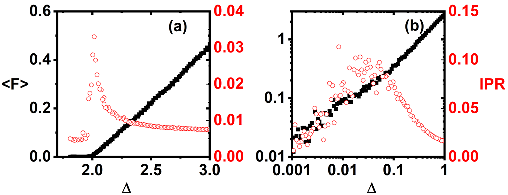}
		\caption{\label{fig4} Synchronization transitions. (a) Uniform frequency distribution where a continuous transition occurs at $\Delta_c=2$. The space averaged velocity fluctuation $\langle F\rangle$ (black) drops linearly toward zero, while the inverse participation ratio $\mathrm{IPR}$ (red) develops a sharp peak exactly at $\Delta_c$. (b) Lorentzian frequency distribution where no true transition occurs. $\langle F\rangle$ smoothly approaches zero only as $\Delta\rightarrow0$; however, the $\mathrm{IPR}$ reaches its maximum at $\Delta\simeq0.06$, beyond which $\langle F\rangle$ increases linearly with $\Delta$.}
	\end{figure}

	When the natural frequencies are drawn from a Lorentzian distribution, genuine quantized synchronous states are practically absent for nonzero $\Delta$. The exponential tails of the distribution guarantee that active oscillators are always present in any large system, precluding exact frequency locking from the start. As a result, the mean fluctuation $\langle F\rangle$ vanishes only in the limit $\Delta\rightarrow0$, consistent with Fig.~\ref{fig4}(b). Nevertheless, both $\langle F\rangle$ and  $\mathrm{IPR}$ reveal a clear crossover near $\Delta=0.06$. Above this value $\langle F\rangle$ grows approximately linearly with $\Delta$, whereas the $\mathrm{IPR}$ peaks at the same $\Delta$, signalling the strongest localization of bursting activity. To mitigate the influence of outliers in the Lorentzian tails, oscillators with extreme natural frequencies ($\omega_i>5$) are excluded when computing $\langle F\rangle$ and $\mathrm{IPR}$.

	\section{Conclusions}
	In this Letter, we examine a ring of oscillators governed exclusively by three-body coupling, the minimal setting in which higher-order interactions arise. We show that even this simplest form of higher-order coupling produces phenomena that have no counterpart in pairwise networks. Our results are four-fold. i) Quantized synchrony without phase coherence: In the absence of disorder, the system supports an extensive number of stable, frequency-locked states with quantized collective velocities $\Omega_n=\sin(2\pi n/N)$, which lack a macroscopic phase order. ii) From discrete to continuous Spectra: Introducing frequency heterogeneity dissolves the sharp quantized levels into broad, overlapping bands of frequency-locked states. iii) Localized bursting: Beyond the critical heterogeneity, global synchrony collapses, giving way to a novel desynchronous state where the bursting dynamics are highly localized. Active oscillators exhibit rapid bursts, while the passive majority remains trapped near the quantized frequency-locked background. iv) An extreme continuous transition for uniformly distributed natural frequencies. The transition at $\Delta_c$ is continuous yet extreme, with the system jumping from complete asynchrony directly to a state of global frequency locking.
	
	These findings extend the theory of synchronization beyond pairwise interactions. Future studies should probe similar dynamics on more complex topologies and with other higher-order interaction kernels, a programme likely to reveal further emergent behaviours. The principles we uncover may also guide the design of real-world multi-body systems, e.g., exploiting quantized frequency-locked states in computing architectures or sensor networks that demand discrete, stable operating modes, and in intelligent autonomous driving, where precise vehicle coordination is essential.
	
	\section{Methods}
	The numerical results reported in this study were obtained through computations performed using FORTRAN.
	
	\section{Data availability}
	The authors declare that the data supporting the findings of this study
	are available within the paper and its Supplemental Material.
	
	\section{Code availability}
	The code used for the current study is available from the corresponding author on reasonable request.

\end{document}